%% file: main.tex
\documentclass[nonacm, sigconf]{acmart}
\renewcommand\footnotetextcopyrightpermission[1]{} 
\usepackage{graphicx}
\usepackage{amsmath}
\usepackage{multirow}
\usepackage{xspace}
\usepackage{adjustbox}
\usepackage{svg}
\usepackage[textsize=tiny,disable]{todonotes}
\usepackage[utf8]{inputenc}
\usepackage{hyperref}
\usepackage{comment}

\usepackage{array}
\newcolumntype{P}[1]{>{\centering\arraybackslash}p{#1}}
\newcolumntype{M}[1]{>{\centering\arraybackslash}m{#1}}

\usepackage[underline=false]{pgf-umlsd}

\newcommand{\removedtodo}[2][]{}

\acmYear{2020}\copyrightyear{2020}
\setcopyright{none}

\begin{document}
%

\title{
Empowering Citizens by a Blockchain-Based Robinson List
}

\titlenote{A preliminary version of this work appeared at CryBlock '20 \cite{cryblock}.}


\author{Albenzio Cirillo}
\affiliation{%
  \institution{Fondazione Ugo Bordoni}
}
\email{acirillo@fub.it}

\author{Vito Dalena}
\affiliation{%
  \institution{Fondazione Ugo Bordoni}
}
\email{vdalena@fub.it }

\author{Antonio Mauro}
\authornote{These authors provided a fundamental contribution in the implementation of the PoC}
\affiliation{%
  \institution{Fondazione Ugo Bordoni}
}
\email{amauro@fub.it }

\author{Francesco Mogavero}
\affiliation{%
  \institution{Fondazione Ugo Bordoni}
}
\email{fmogavero.ext@fub.it}

\author{Diego Pennino}
\affiliation{%
	\institution{Roma Tre University - Eng. Dept.}
}
\email{pennino@ing.uniroma3.it }

\author{Maurizio Pizzonia}
\affiliation{%
	\institution{Roma Tre University - Eng. Dept.}
}
\email{pizzonia@ing.uniroma3.it }

\author{Andrea Vitaletti}
\affiliation{%
  \institution{Sapienza University of Rome}
}
\email{vitaletti@diag.uniroma1.it }

\author{Marco Zecchini}
\authornotemark[2]
\affiliation{%
  \institution{Sapienza University of Rome}
}
\email{zecchini@diag.uniroma1.it }

\renewcommand{\shortauthors}{}


\begin{abstract}
	\input{000_abstract}
\end{abstract}
\maketitle 
\input{010_intro}
\input{020_SOTA}

\input{030_requirements.tex}
\input{040_solution.tex}

\input{050_PoC}
\input{060_evaluation}

\input{065_evaluation1}
\input{070_privacy}
\input{080_conclusions}

\bibliographystyle{unsrt}
\bibliography{RPO}

\end{document}

%% file: 000_abstract.tex

A Robinson list protects phone subscribers against commercial spam calls. Its least basic functionality is to 
collect the denial of the subscribers to be contacted by market operators.
Nowadays, Robinson lists run as centralised services, which implies that citizens should trust third parties for the management of their choices. 

In this paper, we show a design that allows us to realise a Robinson list as a decentralised service.
Our work leverages the experience developed by 
Fondazione Ugo Bordoni as the manager of the Italian Robinson list. 
We present a general solution and a proof-of-concept (PoC) adopting the Algorand technology. 
We evaluate the performances of our PoC in terms of its scalability and of the latency perceived by the involved actors.
We also discuss aspects related to identity management and privacy. 



 

%% file: 010_intro.tex
\section{Introduction}\label{sec:intro}
User contact information is precious gold to marketing strategists. 
For this reason, users are now exposed to an unprecedented number of unwanted calls  on their phones (\emph{spam calls}) for marketing purposes, or even to attempts to defraud them by gaining their confidence (\emph{scam calls}).
To better understand the dimension of this problem, we focus on some of the figures reported in \cite{spamstats} that collects statistics from several sources:
\begin{itemize}
    \item In 2019, roughly 40 percent of all calls in the US were scams. (Source: First Orion)
    \item Americans lost nearly \$19.7 billion from phone scams in 2020 — more than double the amount lost in 2019. (Source: Truecaller)
    \item 44\% of Americans received spam calls related to COVID-19 in Q1 of 2020.(Source Truecaller)
    \item The number of spam calls jumped over 300 percent worldwide between 2017 and 2018. 
 \end{itemize}

One way to tackle the problem of unwanted calls is to block calls on our mobile phone \cite{fcc}. Call blockers have also been recommended by governments in the recent pandemic emergency regarding COVID19~\cite{KoreaCovid}, to prevent scams or other threats for citizens. 
However, such apps will often not distinguish legitimate survey research calls from telemarketing \cite{aapor}. Moreover, call blockers are often eluded by spammers, as they usually have a large set of caller IDs to reach the customer, so blocking a single ID is not effective, most of the times. 

We need a system capable to protect users and, at the same time,  to allow  marketing operators to  reach legitimately a vast audience of informed and aware consumers.
\emph{Robinson Lists} are the standard answer to this need. 
It is an opt-out list of people who do not wish to receive marketing communications. Nowadays, Robinson lists are centralised services existing in a number of countries, such as UK, Canada, Australia, and Italy, just to mention a few.

The Italian Robinson list is called \emph{Registro Pubblico delle Opposizioni} (RPO) \cite{RPO} and it is managed, since 2011, by Fondazione Ugo Bordoni (FUB). 
Today RPO manages $1.5$ millions records, but, with the next introduction, in the near future, of the management of mobile phone numbers, it is expected to be required to handle at least $100$ millions records.
In the work described by this paper, the experience of FUB was fundamental for the  identification of the requirements of a typical Robinson list and of realistic workloads.

\textbf{Motivations.} Nowadays, Robinson lists are managed as  centralised services. The purpose of this work is to design and evaluate a decentralised Robinson list on top of blockchain technologies. 
The main positive effect of this approach is to empower citizens providing them with direct control of their options.
Moreover, institutions that operate Robinson lists should aim to reduce their involvement in management of user options as much as possible, in order to lower their risks and costs.
Indeed, cryptographic techniques adopted by blockchains natively provide tools that easily solve possible controversy between subscribers and operators. In this perspective, in this work we focus on the most critical and urgent function to decentralise, namely the ability of citizen to autonomously opt-in/out.

\textbf{Structure of the paper.} In this paper, we present a solution for a decentralised Robinson list. We first formally describe  requirements on the basis of the experience on the Italian RPO (Section~\ref{sec:requriements}). Then, we describe our solution of blockchain-based Robinson list in abstract terms and identify the   
features that we require from the underlying blockchain technology (Section~\ref{sec:solution}). We recognise that the Algorand technology complies with the identified requirements and base on it our PoC realisation. We describe the architecture of our PoC and some of the most interesting  technical details (Section~\ref{sec:poc}). We exploit our PoC to perform an experimental evaluation of our approach that shows  results that are comparable with the centralised solution currently deployed for the Italian RPO (Section~\ref{sec:evaluation}). We also 
discuss privacy and identity management aspects (Section~\ref{privacy}) and future work (Section~\ref{conclusions}). 

%% file: 020_SOTA.tex
\vspace*{-1cm}
\: \\ \: \\ 
\section{State of the Art}\label{sec:sota}
Robinson lists are traditionally realised by a centralised authority that is
trusted by subjects and operators~\cite{tempest2007robinson}. The scientific
literature regarding systems supporting Robinson lists is very limited. Indeed,
the technical aspects of a centralised Robinson list are not particularly
challenging. 

Blockchain applications for managing user consents have been already considered for other application scenarios (e.g., medical, banking, IoT stream data). 

The Dwarna system described in~\cite{mamo2019dwarna} stores research partners’ consent for bio-banking process in a blockchain to create an immutable audit trail of research partners’ consent changes. The work described in~\cite{bhaskaran2018double} describes the design and implementation of a smart contract for consent-driven data sharing. In this context, service providers can share data about costumers among each other and validate each other’s permissions to share it according to costumers' consent.
Both works are realised on Hyperledger~\cite{androulaki2018hyperledger}, a permissioned blockchain solution. 
The work described in~\cite{rantos2018advocate} proposes, a user-centric solution that allows data subjects to easily control consents regarding access to their personal data in an IoT ecosystem and exercise their rights in accordance with the GDPR regulation. However, a data subject and a data controller cannot publish their digital consents with the corresponding signatures directly on a blockchain but a centralised platform takes care of this task. In our solution, we aim to let each subscriber managing its own consent, autonomously.

Essentially, Robinson lists allow a subject to associate an attribute (the
option) to an identifier (a telephone number). This problem is strictly related
to the one faced by identity management systems (IdM). The main purpose of an IdM is to bind an identifier of a subject (usually a public key) with attributes, claims, entitlements, or properties of that subject. IdMs are standardized by ISO~\cite{ISO-24760} and there regulations about them (e.g., \emph{eIDAS}~\cite{eIDAS}). 
The idea of realising IdM on top of blockchains is a step toward making IdMs independent from a specific organization.  The \emph{Self-Sovereign Identity (SSI)} approach, surveyed here~\cite{MUHLE201880}, envisions solutions in which subjects should be able to create and control their own identity, without relying on any centralised authority. In this context,
public/permissionless blockchains are fundamental tools. W3C has ongoing efforts to
standardize the building blocks of SSI. \emph{Decentralised Identifiers
	(DIDs)}~\cite{w3cDID1.0} are controlled by subjects and possibly securely stored
in blockchains. DIDs are linked with \emph{DID documents} where attributes are listed.
Certain attributes are associated to \emph{Verifiable Claims/Credentials}
(\emph{VC})~\cite{w3cVCDM1.0} which allow the binding between the identifier and the attributes.

%% file: 030_requirements.tex
\section{Modelling a Robinson list: Actors, Use cases, and Requirements}\label{sec:requriements}

In this section, we briefly outline the main actors and functions of the current centralised RPO 
that we consider as a common accepted practice for a Robinson list. Then, we define the main requirements to support a decentralised Robinson list focusing on the decentralisation of the most critic functions for 
citizens empowerment and disintermediation. 

We model a Robinson list as a collection of records 
$\langle \textit{tel}, \textit{opt} \rangle$, where $tel$ is the telephone number of the \emph{subscriber} and $opt$ can assume the values $IN$, if the subscriber opts-in to receive unsolicited calls on $tel$ from \emph{direct marketing operators}, and $OUT$, vice-versa.

The actors interacting with the Robinson list are the following: 

\begin{description}
	\item[Subscribers] A \emph{subscriber} is the owner of one or more telephone numbers and consequently the owner of some records in the Robinson list, where options associated with telephone numbers are recorded. 
 
	\item[Attestator] The \emph{attestator} is a single actor that is responsible for the creation of the Robinson list, for \emph{binding} subscribers to their telephone numbers, and for pruning the phone number lists received from operators (as described below).

	\item[Operators] An \emph{operator} is interested in obtaining the options associated with a list of telephone numbers. Actually, it provides a list of telephone numbers to the organization operating the Robinson list and receives as output a \emph{pruned} list where the numbers of the subscribers who opted-out have been removed. 
	
\end{description}
Therefore, four fundamental functions must be supported by system realizing a functional Robinson list, which are
 listed in the first column of Table~\ref{tab:use-cases}.

\begin{table*}
    \centering\begin{tabular}[c]{|c|c|c|}
\hline 
\textbf{Function}  & \textbf{Centralisation}  
 & \textbf{Actor performing the operation}\tabularnewline
\hline 
 \begin{tabular}[c]{@{}l@{}} Robinson List  Creation \end{tabular} & Centralised & Attestator  \tabularnewline
\hline 
\begin{tabular}[c]{@{}c@{}} Binding of subscribers to phone numbers \end{tabular} & Centralised & Attestator  \tabularnewline
\hline 
 \begin{tabular}[c]{@{}c@{}} Choice of opt-in/opt-out by subscribers\end{tabular} & Decentralised & Subscriber  \tabularnewline
\hline 
Pruning lists requested by operators  & Centralised  & Attestator \tabularnewline
\hline 
\end{tabular}
\caption{Functions and decentralisation choices.}    
\label{tab:use-cases}
\end{table*}

In the current RPO, all these functions are centrally performed by the attestator (i.e., FUB). For opt-in/out and pruning operations, the attestator receives requests from subscribers and  marketing operators, respectively, and acts as a trusted intermediary for accessing the Robinson list. 

\subsection{Towards a Decentralised RPO}

In formally listing the requirements, we primarily focused on fully decentralising opting-in/out operations, giving full control to the subscribers without relying on intermediaries. This guarantees the principle of citizen empowerment that is a motivating argument of our approach. Table~\ref{tab:use-cases} summarises our choices regarding decentralisation.
 


In the current proof-of-concept (described in Section~\ref{sec:poc}), the other functions are left centralised leaving the investigation of their possible decentralisation as future work. This choice is also motivated by the following considerations.

We do believe that it is practical and acceptable that the creation of the assets in the Robinson list is centralised, provided that their management (i.e. opt-in/opt-out choice) is fully decentralised.  
To the best of our knowledge, centralised certification is the only legal method that is currently accepted by Public Bodies to bind subscribers to their phone numbers. We further discuss this problem and possible approaches in Section~\ref{privacy}.


About the decentralisation of the pruning function, we consider it as an important part of our future research work 
(see Section~\ref{conclusions}).



The following is a formal list of the requirements of our decentralised Robinson list.
\begin{enumerate}
	
\item \label{req:records} Each subscriber can own any number of records $\langle
\textit{tel}, \textit{opt} \rangle$ (for distinct values of \textit{tel}).

\item \label{req:ownership} 
The owner of a record should be able to switch the $\textit{opt}$ value.

\item \label{req:query} An operator should be able to query the list for the options 
related to specific values of \textit{tel}.

\todo[inline]{Pzz:maybe.Fra: Commentare \ref{req:proof} dato che non è stato implementato? In questo modo risparmiamo 35 parole ed evitiamo richieste di spiegazioni sull'implementazione di questo punto da parte dei reviewers. \\
Poco più sotto riprendiamo \ref{req:proof} accostandolo allo stato della blockchain (ma senza dire nulla sulla prova); commentando anche quella parte risparmieremmo altre 133 parole}
\item \label{req:proof} There should be the possibility to equip the result of a query (or of a pruning operation) with a succinct cryptographic 
proof, to be shown to anyone, that the result of the query is correct. 

\item \label{req:decentralisation} No single actor (comprising the attestator) or
small colluded set of actors, should be able to independently change records
belonging to others without the consent of the owner. 

\end{enumerate}

It is worth noting that Requirements~\ref{req:records},~\ref{req:ownership}
and~\ref{req:query} are also met by centralised Robinson lists requiring trust in a central authority. Vice versa, Requirements~\ref{req:proof} and~\ref{req:decentralisation} 
 are normally not met by
current centralised approaches.

Requirement~\ref{req:decentralisation} naturally lead the design toward a
blockchain-based approach. However, a simple solution may not met
Requirement~\ref{req:proof}. Consider for example the trivial solution in which
two special addresses IN and OUT are present and the a subscriber opted for IN
or OUT by performing a transaction on the respective address. A proof of the
choice of a subscriber at a certain time requires to analyse the history of the
blockchain back in time, up to the last choice performed by the subscriber.
This is not a succinct proof.

Essentially, the problem is that the options are not encoded in the current
state of the blockchain. In Section~\ref{sec:solution}, we describe a solution
that does encode options in the blockchain state. This make fulfilling
of Requirement~\ref{req:proof} much easier to implement, especially on technologies that explicitly store the states of the accounts.




%% file: 040_solution.tex

\section{Solution}\label{sec:solution} 

Motivated by Requirements~\ref{req:ownership},~\ref{req:proof}
and~\ref{req:decentralisation}, we realise Robinson List on a blockchain
technology. There are a large number of different blockchains, showing
different trade-offs between supported features and scalability. In devising our
solution, we aimed at requiring a blockchain with a set of features that allow us to fulfill the requirements introduced in Section~\ref{sec:requriements}, and supporting high scalability, in view of the future management of a large number of records (about 100Mln considering the italian RPO alone).
In addition to standard blockchain
characteristics
, we require the blockchain technology to support the following features: 

\begin{itemize}

\item The blockchain should support the creation and transfer of custom \emph{tokens} to represents \emph{IN} and \emph{OUT} options made by subscribers.

\item  At each instant in time, the choice of a subscriber is either \emph{IN} or \emph{OUT}.  As will be clear in the following,  we need to atomicly swap a token IN with a token OUT. For this reason the blockchain should support grouping of  transactions so that the whole group is either committed or not. 


\item The blockchain should support \emph{smart contracts} to manage citizen choices by an algorithmic governance. In this paper, we are interested in proving that even stateless smart contracts -- the less demanding ones -- are enough for our purposes. 


\end{itemize}


Custom tokens are supported on many blockchains, either natively~\cite{asset} or by using proper smart contracts (see for example~\cite{victor2019measuringErc20}). In our solution, tokens are used to store the current opt-in or opt-out of a subscriber without relying on
smart contracts with persistent state.



In section~\ref{sec:poc}, we present our prototype based on the Algorand
technology, which provides all these features. 

We now describe our solution. Subscribers are identified by one or more public keys and each
telephone number is associated to a public key. How this
association can be carried out is discussed in Section~\ref{privacy}. Private
keys are secretly held by the corresponding subscriber. The attestator is also
associated with a key pair.
The binding between the public key $p_k$ and the telephone number $tel$ is stored on chain. We decided to encrypt those element so that only the key's owner and the attestator can view the values in clear text. This decision was made to preserve citizen's privacy, by masking the binding of his/her private key to the phone number.

The binding information, namely the pair $(p_k, tel)$, is hidden into the payload of a transaction that the attestator sends to the user. The masking technique consists in the following steps:
\begin{enumerate}
\item the attestator generates $r_k$, a random 128 bit AES symmetric key and a random 128 bit initialisation vector $iv_k$;
\item The attestator uses $r_k$ and $iv_k$ to encrypt the pair $(p_k, tel)$ by using AES-CBC (Cipher-Block-Chaining) scheme, obtaining $eb_k$;
\item $r_k || iv_k$ 
is encrypted with user's public key and attestator's public key,  
producing the values $r_{ku}$ and $r_{ka}$, respectively (these encryptions are performed
using the PyNaCl library~\cite{PynaclDocu, NaClPaper} and the X25519 elliptic curve~\cite{rfc7748}); 
\item The triple $\langle eb_k, r_{ku}, r_{ka} \rangle$ is written on-chain.
\end{enumerate}

Smart contracts are associated with an identifier that, for many aspects, plays the
same role of the public key. Hence, for homogeneity, we call it the public key
of the smart contract and, when we generically refer to public keys, we intend
also to include smart contracts identifiers.

In our solution, we define two custom class of tokens named \emph{IN}  and
\emph{OUT}. We call IN tokens and OUT tokens the individual units of IN and OUT,
respectively.  In the system, the global number of IN tokens and of OUT tokens
is always the same. To streamline description, we refer to IN as the
\emph{opposite} of OUT and vice versa.

Each public key is associated with a \emph{wallet} that holds tokens. Public
keys are used to identify source and destination wallets in transactions
involving tokens. In our solution,  the tokens are kept in smart contract
wallets. For simplicity, we also say that the smart contract \emph{contains} or
\emph{stores} the tokens.

Each telephone number $t$ in the Robinson list 
is associated with a corresponding smart contract
$U_t$. The system is built to enforce the following \emph{option constraint}:
for each telephone numbers $t$, the wallet $U_t$ contains either only one IN
token or only one OUT token. $U_t$ has been designed to avoid the exchange of tokens between subscribers that can lead to the violation of the option constraint. 
We can express three possible states
for each telephone number: opted in, opted out, or no option expressed. The third
state is represented by the absence of $t$ and $U_t$ in the system. In our approach there exists another
smart contract $C$ that is in charge of realizing the \emph{switch} of the token
stored in $U_t$ with its opposite, while fulfilling the option
constraint. It also stores all tokens that are not currently stored by any $U_t$ contract
and that may be needed in future switch operations. 

Intuitively, a switch operation is performed as follows. Consider a telephone
number $t$ with smart contract $U_t$ containing an $O$ token (where $O$ is
either IN or OUT). The switch operation consists of the following two transfers:
(1) transfer the $O$ token currently in $U_t$ to $C$ and (2) transfer an
$\overline O$ token (where $\overline O$ is the opposite of $O$) from $C$ to
$U_t$. The above transfers are asked by the subscriber of $t$ as a group to be
atomically executed. Smart contracts perform the following checks:  $U_t$ checks that group is signed 
by the private key of $t$ (i.e., it come from the owner of $t$),
 $C$ checks that the two transfers are
consistent with respect to the option constraint. Note that, for security reason,  all $U_t$'s should conform to a
template of smart contract in which the only changed part is related to the
public key of $t$ that the contract use to verify the source of the accepted
transactions. We call this the \emph{standard template}.

We now formally describe the steps involved in the  use cases
that initialise or change the content of our Robinson list.

\paragraph{Robinson list initialisation.} To initialise our decentralised Robinson list
the following operations are performed.
\begin{enumerate}

\item The attestator generates its public/private key pair.	
	
\item The attestator creates IN and OUT tokens in a quantity which is supposed
to be much larger than the amount of telephone numbers to be managed.

\item The attestator deploys a smart contract $C$.

\item The attestator transfers all IN tokens to $C$.

\end{enumerate}

\paragraph{Addition of a new telephone number.} To add a new telephone number $t$ to
the Robinson list, the following operations are performed.

\begin{enumerate}

\item The subscriber of $t$ creates a new public key $k_t$ and the corresponding
private key. It also creates, the corresponding $U_t$ according to the standard template.

\item The subscriber asks the attestator to associate $t$ with public key $k_t$
(see Section~\ref{privacy}) and to add $t$ to the Robinson list.
 
\item The attestator verifies the association of $k_t$ with $t$ (see
Section~\ref{privacy}) and that $U_t$ conforms to the standard template. If
checks are successful, the attestator sends one OUT token to $U_t$. This represents a
default opt-out state for a newly listed telephone number.

\end{enumerate}

\paragraph{Option switching.} To switch the option for a telephone number $t$ whose smart contract
$U_t$ contains an $O$ token, the following operations are performed. 

\begin{enumerate}

\item The subscriber of $t$ prepares two transaction $\tau$ and $\tau'$ as follows:
   \begin{itemize}
   	\item $\tau$ transfers one unit of $O$ from $U_t$ to $C$, and
   	\item $\tau'$ transfers one unit
   	of $\overline O$ from $C$ to $U_t$.
   \end{itemize}
   
\item The subscriber bundles $\tau$ and $\tau'$ into a group to be atomically
executed, signs the whole group with the private key associated with $t$ and
broadcasts it.

\item Within the execution of the consensus algorithm, both $C$ and $U_t$ are
executed: 
	\begin{itemize}
		
		\item $C$ checks that the two transfers are to and from the same $U_t$ and
		that transfers are for single opposite tokens (to force the option constraint),
		
		\item $U_t$ checks that the request comes from $t$.
	\end{itemize}
If checks are successful and regular balance constraints are fulfilled, both
transactions are committed.

\end{enumerate}


%% file: 050_PoC.tex
\section{Proof of Concept}\label{sec:poc}

In this section we present the implementation of a proof-of-concept (PoC)~\cite{poc} of the decentralised Robinson List solution described in Section~\ref{sec:solution}. The main goal of the PoC is to show how the proposed solution can be implemented on a blockchain technology and to provide a first evaluation on the performance of this implementation.


\subsection{Technical components}

The PoC has been developed on Algorand \cite{algorand} because it supports all the necessary features to implemented our solution as  discussed in Section~\ref{sec:solution} and it is one of the solutions claiming to address the blockchain trilemma \cite{ethfaqsharding, 10.1145/3410699.3413800}, thus  providing state-of-the-art performance in terms of latency and scalability.

Its inter-block time is very short: about 5 seconds. 

To support the reader in a better understanding of our PoC, we briefly describe how Algorand implements smart contracts, tokens management and atomically committed transactions.



\emph{Algorand Standard Assets} (ASA) are standard mechanisms for creating, managing, transferring and destroying \emph{digital tokens} (or \emph{assets}). 
In Algorand, an account should explicitly allow their use to be able to receive them.


\emph{Algorand Smart Contracts} (ASC) are small programs that serve various functions on the blockchain and operate at layer-1. Smart contracts are separated into two main categories, \emph{stateless} and \emph{stateful}. The language for coding ASC is named \emph{Transaction Execution Approval Language} (\emph{TEAL}). Stateless ASC can validates and signs transactions and it can be used as a \emph{Contract Account}. A Contract Account looks like any 
end-user account except that it validates spending transactions according to its code logic. 

Recently, Algorand released stateful ASC that provide local variables. These variables could be used to store the citizens choices thus providing an alternative way of implementing our solution. One contribution of our PoC is to prove that even simple stateless ASC are enough for our goal.

\emph{Atomic Transfers} are indivisible and irreducible batch operations where a group of transactions are submitted as a unit and all transactions in the batch either pass or fail. In our PoC, Atomic Transfers handled by a suitable ASC are employed to swap an $IN$ asset for an $OUT$ one without the involvement of any intermediary and vice versa.



\subsection{Architecture}
\label{sec:architecture}
Now we formally describe the architecture of our PoC.



There is one attestator and several subscribers, each one with only one telephone number associated with its public key. 
This association is kept encrypted in the blockchain (see Section~\ref{sec:solution}) and cached in a database by the attestator. 
In the current implementation, the $U_t$ contract has not been implemented  and the IN/OUT token is kept directly in the subscriber account. We recall that the main goal of the PoC is to evaluate feasibility and performance, and the $U_t$ contract, whose main purpose is to  achieve a higher level of security, has a very minor impact on that and it will be consequently implemented in future releases.

\begin{figure}
	\centering
	\includegraphics[width=\linewidth]{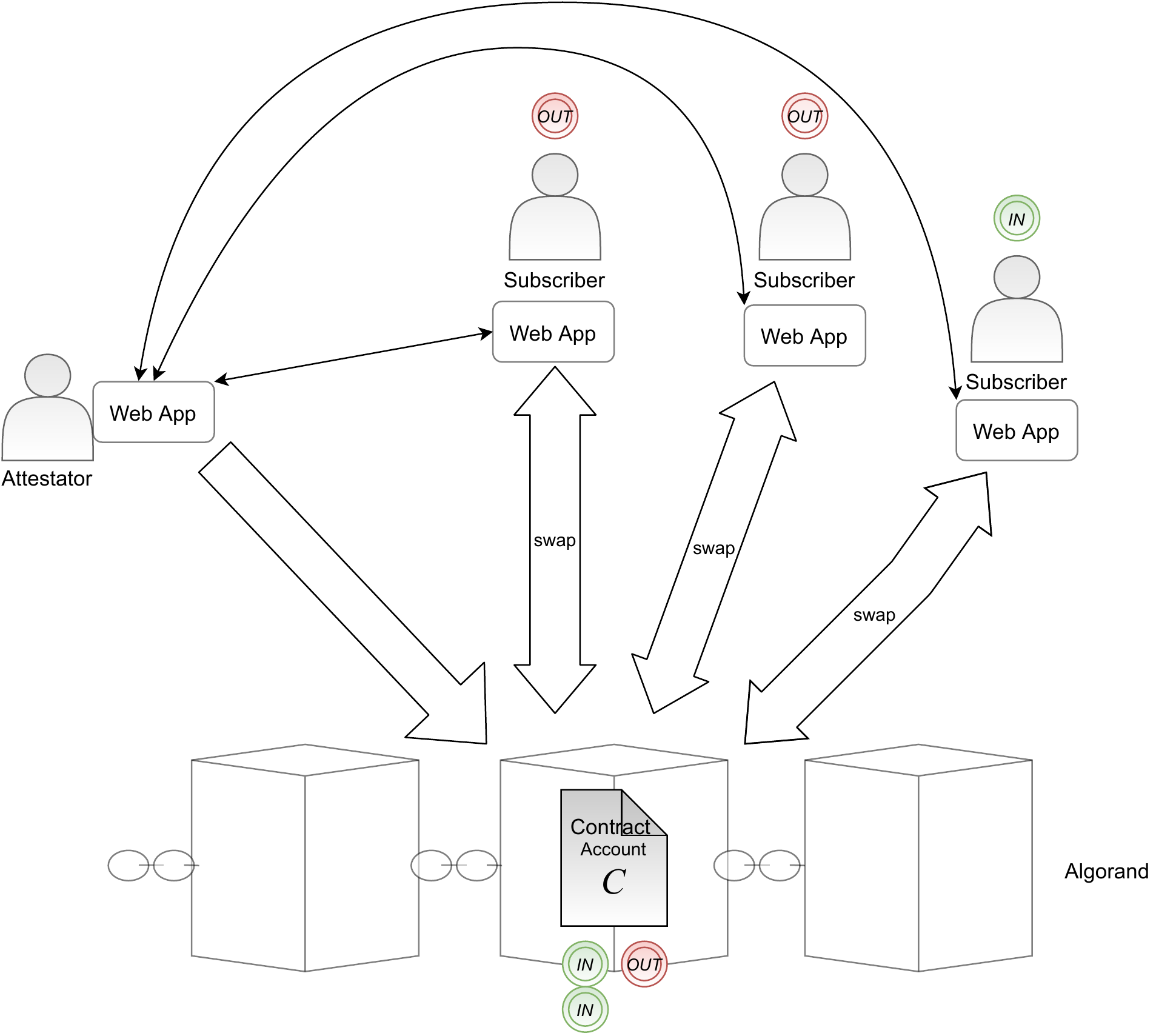}
	\caption{Architecture of the proof-of-concept}
	\label{fig:poc}
	\vspace{-0.5cm}
\end{figure}

Figure~\ref{fig:poc} shows a simplified architecture of the system.
The whole interaction with the Algorand blockchain back-end is performed by a web application developed with the Django framework.
 
The attestator set-ups the system generating all the $IN$ and $OUT$ assets. Then, it deploys the Contract Account $C$ and explicitly allow $C$ to receive both $IN$ and $OUT$ tokens. The set-up ends with an asset transfer transaction of all $IN$ tokens from the attestator to $C$. 

During the initialisation of the subscriber, 
the attestator binds the public key of the subscriber to its mobile phone (see Sections~\ref{sec:solution} and~\ref{privacy}). 
The subscriber explicitly enables reception of $IN$ and $OUT$ tokens in its wallet, then, if the binding is successful, the attestator transfers one $OUT$ token to the subscriber wallet. For the sake of simplicity, current PoC assumes that  binding is  always successful. The initialisation is then completed transferring one $OUT$ token from $C$ to the subscriber wallet. This is done at the first access of the user into the web interface.



The subscriber can swap the $OUT$ token with an $IN$ one by executing an Atomic Transfer with $C$. The TEAL code of $C$ is designed to approve the atomic transfer only if it consists of two transactions: (a) $OUT$ token from the subscriber's wallet to $C$ and (b) $IN$ token from $C$ to the subscriber's wallet. A symmetric atomic transaction lets the user swap the $IN$ token with the $OUT$ one. \\

%% file: 060_evaluation.tex
\section{Evaluation}\label{sec:evaluation}
The proposed solution is functionally effective, as it proves that a users can express their choice of being contacted on their personal phone numbers on a distributed and decentralised ledger. 

Nevertheless, in order to be actually employed, the solution should be comparable to the current centralised one in terms of costs and performance.
 In fact, while a blockchain solution provides an evident improvement in terms of resilience to network failures or data integrity and traceability, it might suffer of significant issues in terms of scalability, bandwidth, latency and operational costs, in particular in view of a possible public/permissionless deployment.
 
 The purpose of this section is to quantitatively measure to what extent our approach can actually successfully compete with the  current Italian centralised solution for the italian \emph{RPO} service (see Section~\ref{sec:intro}).
 Our testing activity was carried on  running our PoC on a VM on a linux OS with 40GB of disk and 4GB of RAM. 

The environment has been implemented on docker and runs in three containers.
\begin{description}
\item[Web.] The presentation layer.
\item[DB.] The identity database (owned by the attestator) that caches the binding between the public keys (i.e., the blockchain addresses) and the telephone number of the subscriber.
\item[Algorand-Node.] The Algorand node realising a private network.
\end{description}
We evaluated the performance of our prototype in terms of time consumption, investigating the different phases whose subscription process is made of. 
In the decentralised scenario, the subscription event is composed of 5 different phases:
\begin{itemize}
\item \textbf{Create Wallet} - subscriber's wallet, that should contain the OUT or IN token, is created;
\item \textbf{Init Wallet} -  the attestator transfers a minimal amount of coins to the wallet, to enable the subscriber to perform transactions on the blockchain;
\item \textbf{Binding} - the attestator binds the public key of the subscriber to its mobile phone writing it on the blockchain in an encrypted form (as described in Section~\ref{sec:solution}) and in clear text in the local identity database;
\item \textbf{Opt-In ASA} - the subscriber explicitly enables reception of $IN$ and $OUT$ tokens in
its wallet;
\item \textbf{Transfer ASA} - the attestator transfers the $OUT$ token to the subscriber's wallet.
\end{itemize}

%% file: 065_evaluation1.tex





RPO currently manages over $1.5$ million of users' fixed phone numbers and, starting from 2021, should be able to support over $100$ million of users' phone numbers, as also mobile numbers will be included. 
 
Thanks to the data provided by FUB, responsible for managing RPO's infrastructure, we were able to analyse the history of users' interaction since 2011 (when RPO was initially deployed), in order to obtain aggregated data about subscription rate.
A first concern comes from the fact that a public service, like the one offered by RPO, can suffer from peaks of legitimate service requests known as \emph{flash crowding attack}. This particular event can occur as a consequence of mass announcements (e.g. through TV or journal advertisements), that are usually followed by users concentrating their requests in a very short time interval straight after the announcement, resulting in something very similar to a DDoS attack.

We observed this situation in RPO's history, finding few dates when more than $20000$ subscriptions occurred in a day, which is an extremely high number with respect to the ordinary subscription daily rate.

In order to provide a complete analysis of our proposed architecture, we used this particular events as a scalability benchmark.

\begin{figure}
	\centering
	\includegraphics[width=\linewidth]{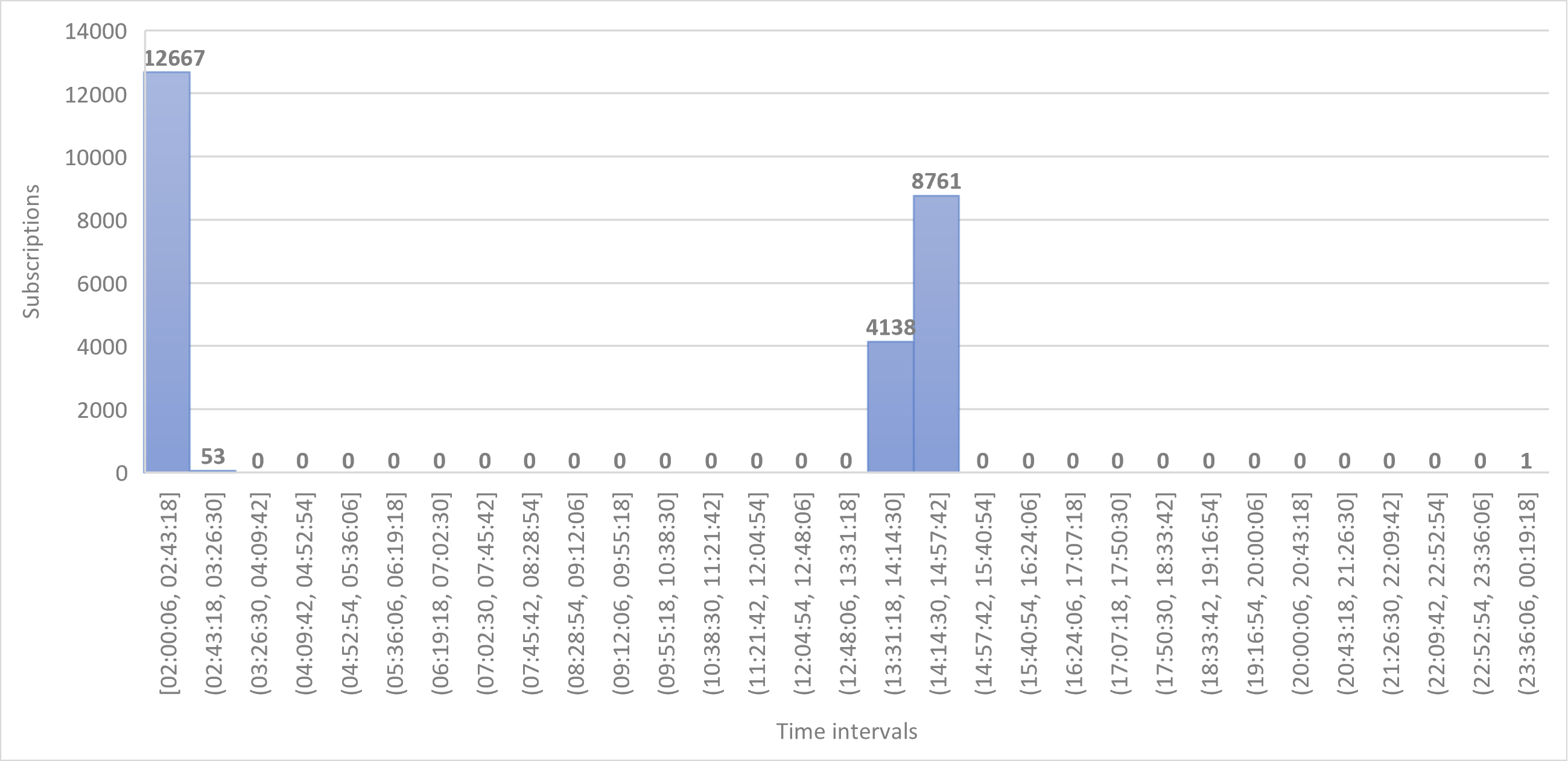}
	\caption{A flash crowding day (2nd Feb 2011) }
	\label{fig:RPOhistogram}
	\vspace{-0.5cm}
\end{figure}

The histogram of the subscriptions, namely the insert operation of a new phone user on a centralised database, during a flash crowding day (in 2011) is represented in figure \ref{fig:RPOhistogram}. 

We observed a total of $25566$ subscriptions in a day, grouped into two intervals in the day. The gathering of subscriptions in two intervals of the day (one at night time and the other around noon) confirmed us that the \emph{insert on database} operations were executed in batch scripts. 

Analysing our log in peak periods we calculate that the average time for the insert operation is about $200 ms$.

We decided to use standalone accounts rather than wallets to overtake the operational cost of creating a wallet.
A \emph{standalone account} is an Algorand address and private key pair that is not stored on disk. 
A user can invoke, on his/her premises, the algo-sdk to generate a pair of public and private keys. The time cost for generating an account is significantly reduced. With reference to the benchmark, we evaluated the time consumption for registering 20’000 users:
\begin{table}[h!]
    \centering
    \begin{tabular}{|l|c|}
      \hline
      20’000 users subscription completed in    & 00:09:10.709 \\
      \hline
      average time for a single subscription & 00:00:00.267 \\
      \hline
      dev std time for a single subscription & 00:00:00.106 \\
      \hline
    \end{tabular}
    \caption{Time consumption for registering 20’000 users}
    \label{tab:time_20kusers}
\end{table}
In this way we reached 3.6 subscriptions per second, which is a comparable rate with the one observed during flash crowding events.

\subsection{Pruning Test}
\label{sec:pruningtest}
Currently, the main service offered by the current centralised RPO consists in filtering a list of telephone numbers provided by a marketing operator, in order to return them the same list but without the opted-out subscribers. This operation is also known as pruning.
We performed a pruning test operation on a list of 10'000 mobile phone numbers. Indeed, 10'000 numbers is the mean amount of phone numbers for which the telemarketing operators request the pruning activity to RPO.\newline 
Our decentralised RPO can operate the filtering process in two different modes:
\begin{itemize}
    \item by looking on chain for the opt status associated with a specific public key bound to a phone number;
    \item by enquiring a private database that acts as a cache for the binding information on chain, 
     similarly to what happens in the current centralised implementation. 
\end{itemize}

The former solution is preferable because it would prevent RPO to maintain a private database with sensitive information, lowering the risks in case of malicious attacks. Nevertheless, as the binding between the public key and the phone number has been encrypted on the chain, and considering that the phone number is the only input to the filtering process, the first solution would also require to exhaustively look into the whole blockchain, decrypting the binding transactions, till finding the desired phone number and associated public key, which is evidently more expensive than performing a search over a private indexed database. 

The difference in cost between the two methods has been evaluated by comparing 3 different pruning tests, producing the following results: 
\begin{table}[h!t]
\centering
\begin{tabular}{|c|c|c|}
\hline Iteration & Blockchain only & Cache DB  \\
1 & 1'59" & 14" \\
2 & 1'54" & 16"  \\
3 & 1'58" & 16" \\
\hline
\end{tabular}
\label{tab:timepruning}
\caption{Pruning Blockchain only and Cache DB}
\end{table}
As expected, the pruning operation performed without a cache DB, which means using only on chain data, requires a higher time effort.
Considering $B$ as the total number of blocks and $m$ the number of mobile phones to be pruned, in the worst case the cost of search is $O(B*m)$, assuming $B >> m$.
This result suggests for further investigations in order to foster the usage of on chain data in real world processes and not only for notarization purposes, while preserving security and privacy aspects.

\subsection{Dimensioning archival nodes}
\label{sec:nodedimension}
\begin{figure}
	\centering
	\includegraphics[width=\linewidth]{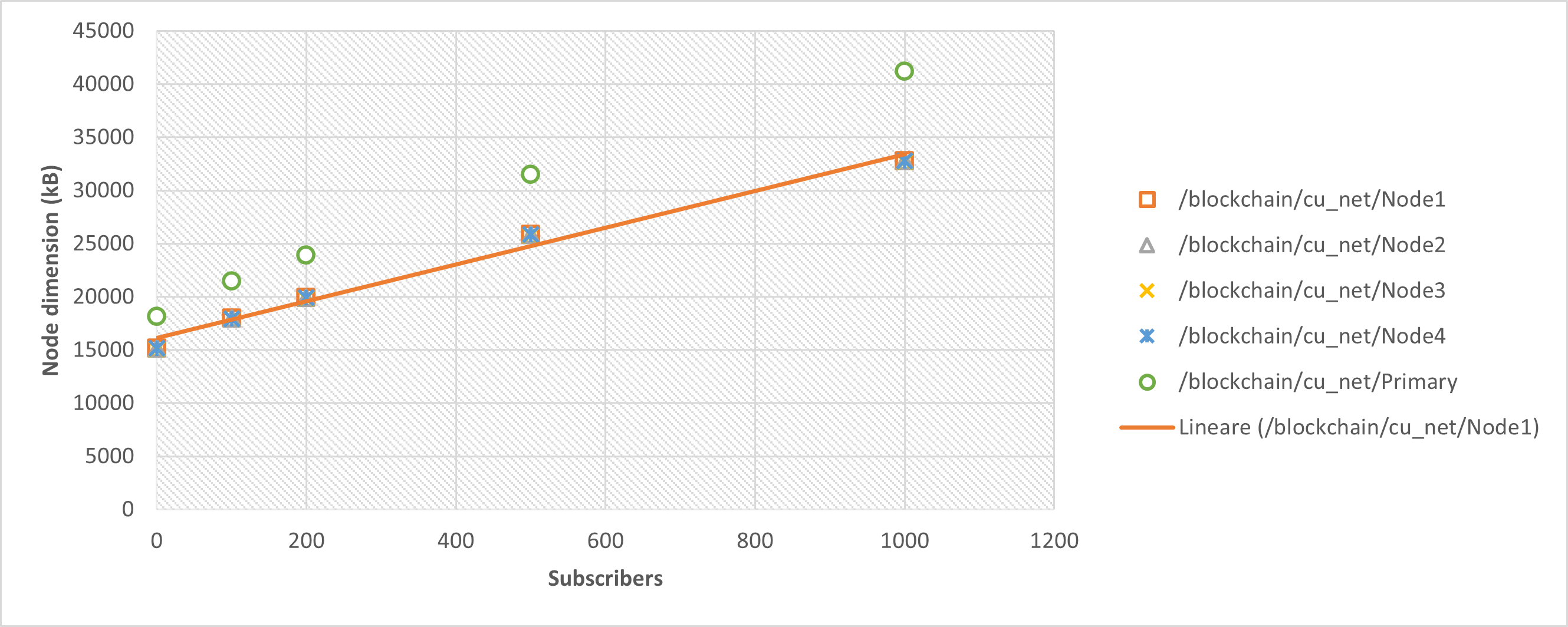}
	\caption{Archive dimension of a node in relation to the number of subscribers}
	\label{fig:nodedimension}
	\vspace{-0.5cm}
\end{figure}

We investigated the storage requirements for archival nodes in a private Algorand blockchain, as the volume of traffic and the dimension of the transactions can vary significantly with respect to the main public network.
In our test, we measured the amount of disk space occupied on 5 different archival nodes when increasing the number of subscribers to the decentralised robinson list.
Results are shown in figure \ref{fig:nodedimension}.
All nodes occupy the same amount of space when increasing the number of subscribers to the service, except for the Primary Node (circle dots) which acts as a relay node and needs to store log activities.

Giving this linear dependence among number of suscribers and disk occupation, we
can estimate the dimension for an archival node that should be able to support a
desired number of subscribers: in our case, we estimated the need for $1650 GB$
for a node that should contain 100 million of subscribers. On top of that
estimate, a blockchain designer should consider the estimate number of
transactions that could be performed, so to define a solid configuration for the
archival node.

Unfortunately, this considerations are not sufficient to dimension an archival node on a private blockchain. Indeed, the blockchain itself generates \textit{empty blocks} 
whenever there are no transactions to be committed. This is typical for all the blockchains and is needed to ensure integrity and security.
On a public blockchain it is unlikely to observe empty blocks, however, in private blockchain dedicated to specific regional services (which might be our case), idle times might be common, especially at night.


\subsection{Economic costs} 


In this section, we analyze the economic  
costs of our approach, focusing on our PoC realization. 

The following are the main costs a subject has to sustain in the proposed decentralised solution: 

\begin{itemize}
    \item Each subscriber needs to possess a minimum amount of Algos to participate to the network. In our PoC, the attestator supports this cost transferring to each subscriber these Algos.
    \item A subscriber opts-in to trade both IN and OUT tokens, issuing two transactions.
    \item The attestator transfers 1 OUT token to each subscriber.
    \item Every time a subscriber wants to change state, it publishes an atomic transfer composed by two transactions to swap the two different tokens. The first transaction is in charge of the subscriber, the second one is paid by $C$. The balance of $C$ is monitored by the attestator that refills it with Algos, when necessary.
\end{itemize}

In Algorand, each transaction costs 0.001 Algos. Moreover, each account must maintain a minimum balance to be considered active (i.e. use an account in transactions) and an additional balance is needed to trade ASAs (i.e. use an account in transactions involving the ASA). The minimum balance required to activate an account is 0.1 Algos~\cite{account} and the one to trade a token is 0.1 Algos~\cite{asset} per ASA. Since, in our scenario, each subscriber trades two ASAs, the overall minimum balance is 0.3 Algos.

Considering the current 1.5M RPO subscribers, the attestator pays 450000 Algos for the minimum balance of the subscribers, 1500 Algos to transfer it and 1500 Algos to distribute the OUT tokens. Hence, the attestator pays 453000 Algos.

Each subscriber pays 0.002 Algos to publish the transaction opting-in the IN and OUT tokens. Swapping the ASA costs 0.002 Algos in total. Hence, the subscriber pays 0.003 Algos. 



The cost of the initialisation phase is relevant with respect to all other
costs. Supposing to use the public Algorand main network, at the current
quotation of the Algo cryptocurrency, the cost of initialising the system for
1.5M subscribers is about 500k€. While this may seems very high, it is
substantially less than the current \emph{annual} costs paid by market operators
to use RPO. In fact, on average, RPO filters about 400 million of numbers each
year. As by May 2021, operators pay, for each filtered number, from 0.004€ to
0.025€, depending on the subscribed bundle. Hence, market
operators collectively pays \emph{annually} more than 1.6M€, which are supposed
to cover just operational costs, since RPO is a not-for-profit service. One
aspect to consider in this discussion is that our approach does not fully substitute
the current centralised realization, hence, some of the costs of the current solution are
supposed to exists also in our approach, like identity verification costs. 
However, we think that this simple analysis is enough to say that in case of realization on the public blockchain 
costs are still comparable with the current centralised approach. 




%% file: 070_privacy.tex

\section{Privacy and Identity Management}\label{privacy}



As already observed, the main contribution of this paper is the decentralisation of the opt-in/opt-out choice. Subscribers choices are \emph{personal data} that have to comply with  \emph{privacy regulation}, like for example the GDPR~\cite{voigt2017eu}. Decoupling the data from the personal identity (\emph{pseudonymization}) is a widely used approach for this kind of compliance. In the previous sections, we deliberately avoided to introduce any in-chain information
that can help to associate choices to related telephone numbers or subscribers.
From this point of view, the solution we have described so far is not affected
by privacy concerns. However, in practice, operators do need to know both
telephone numbers and choices. In this case, regulations require that operators
can have knowledge of that binding, but no other subject should be able to get
it. 

Creating and storing bindings between telephone numbers and public keys are
essentially  \emph{identity management} problems~\cite{ISO-24760}. The simplest
approach to them is to delegate both verification of this binding and the
management of the corresponding database to a trusted third party. For example,
the identity database could be managed by the same attestator $A$ introduced 
in Section~\ref{sec:solution} (and this is the choice in our PoC). When a
subscriber $S$ asks $A$ to add to the Robinson list a telephone number $t$, with public key
$k_t$, $A$ should check the binding and store it in its
identity database. For example, $A$ can ask the subscriber to sign a random
challenge $c$ using the private key paired with $k_t$, where $c$ is
communicated to $S$ by making use of $t$ (e.g., by SMS or by voice call). 
In pruning operations, operators should ask $A$ to obtain
either $k_t$ or directly the current choices for $t$. 

Our solution, with this identity management approach, might be deemed acceptable and regarded as a
substantial improvement with respect to current practices. 
\todo{Pzz:maybe. Fra: Secondo me dovremmo motivare perché è accettabile e perché (ed in cosa) migliora le current practises.
Siccome in 030\_requirements abbiamo detto: "centralised certification is the only legal method that is currently accepted by Public Bodies to bind subscribers to their phone numbers. We further discuss this problem and possible approaches in Section~\ref{privacy}.", potremmo ripetere lo stesso concetto qui. Ad esempio:
Our identity management approach is acceptable if we consider centralised certification is the only legal method that is currently accepted by Public Bodies to bind subscribers to their phone numbers.}
The obvious next step 
is to ditch any involvement of $A$ from regular operation of the Robinson list.

In the just described scenario, $A$ performs three tasks that we would like to 
decentralise. 
\begin{enumerate}
	\item~\label{tsk:check} It checks bindings of telephone numbers to public keys. 
	\item~\label{tsk:store} It stores them into a private identity management database.
	\item~\label{tsk:reply} It replies to operators queries.
\end{enumerate}
Task~\ref{tsk:store} is easy to decentralise, since the \emph{identity pairs}
$\langle t, k_t\rangle$ can easily be stored in clear text into the blockchain. However,
this would be a clear privacy violation, since
blockchain is supposed to be accessible to a multitude of subjects. If we accept to keep
Task~\ref{tsk:reply} centralised, then the identity pairs can be 
encrypted 
so that only $A$
and the owner of $t$ can decrypt them, as described in Section~\ref{sec:solution}. 
A scheme that allows operators to access
identity pairs autonomously, while respecting privacy regulation, is more 
challenging to devise. This is especially
true if we admit operators to be granted or denied access, dynamically. By the way, this
also introduces the problem of who is authorized to grant or deny access to operators or
if this should be performed in a decentralised manner, as well. We intend to develop
these aspects in the future.

Regarding Task~\ref{tsk:check}, the work in~\cite{binding} describes two
blockchain-based approaches to perform this kind of checks in a decentralised fashion.
It relies on a randomly selected committee of participants to the blockchain
that is different for each check and it is hard to predict in advance. Each
member performs the check autonomously and then write in the
blockchain its ``proof'' about the binding. A summary of the proofs can then
be computed at the consensus level and written on-chain as a regular identity pair.
Again, while this approach looks promising,
privacy aspects are yet to be developed.

%% file: 080_conclusions.tex
\section{Conclusions and Future Work}\label{conclusions} 

In this paper, we show the technical feasibility of a decentralised Robinson
list and the adequacy of the performances of our approach. Our solution enables
citizens to express their choice in complete independence while costs, even in
case of adoption of a public blockchain, are comparable with costs of current
centralised RPO. The validity of our approach was also recognized in public
competitions\footnote{See Future of Blockchain 2 competition
\url{https://medium.com/future-of-blockchain-competition/future-of-blockchain-2-summary-and-prizes-e87f3c6f392f}}. 
Concerning future works, we plan to investigate a completely decentralised
solution (including both pruning and binding) where we see two main challenges: conformity to privacy regulation and automatic management of fees, where citizens may possibly be rewarded for
accepting marketing calls.

%% file: main.bbl
\begin{thebibliography}{10}

\bibitem{cryblock}
Albenzio Cirillo, Antonio Mauro, Diego Pennino, Maurizio Pizzonia, Andrea
  Vitaletti, and Marco Zecchini.
\newblock Decentralized robinson list.
\newblock In {\em Proceedings of the 3rd Workshop on Cryptocurrencies and
  Blockchains for Distributed Systems}, CryBlock '20, page 1–6, New York, NY,
  USA, 2020. Association for Computing Machinery.

\bibitem{spamstats}
35+ phone spam statistics for 2017, 2018, 2019 n.
\newblock
  \url{https://www.comparitech.com/blog/information-security/phone-spam-statistics/},
  Last accessed April 2020.

\bibitem{fcc}
{Federal Communications Commission}stop unwanted robocalls and texts.
\newblock
  \url{https://www.fcc.gov/consumers/guides/stop-unwanted-robocalls-and-texts},
  Last accessed February 2020.

\bibitem{KoreaCovid}
{ROK Ministry of Economy and Finance}flattening the curve on covid-19.
\newblock
  \url{http://www.undp.org/content/seoul_policy_center/en/home/presscenter/articles/2019/flattening-the-curve-on-covid-19.html},Last
  accessed April 2020.

\bibitem{aapor}
American~Association for Public Opinion~Research.
\newblock Spam flagging and call blocking and its impact on survey research,
  February 2018.

\bibitem{RPO}
Registro pubblico delle opposizioni.
\newblock \url{http://www.registrodelleopposizioni.it/en}, Last accessed April
  2020.

\bibitem{tempest2007robinson}
Alastair Tempest.
\newblock Robinson lists for efficient direct marketing.
\newblock In {\em International Direct Marketing}, pages 129--152. Springer,
  2007.

\bibitem{mamo2019dwarna}
Nicholas Mamo, Gillian~M Martin, Maria Desira, Bridget Ellul, and Jean-Paul
  Ebejer.
\newblock Dwarna: a blockchain solution for dynamic consent in biobanking.
\newblock {\em European Journal of Human Genetics}, pages 1--18, 2019.

\bibitem{bhaskaran2018double}
Kumar Bhaskaran, Peter Ilfrich, Dain Liffman, Christian Vecchiola, Praveen
  Jayachandran, Apurva Kumar, Fabian Lim, Karthik Nandakumar, Zhengquan Qin,
  Venkatraman Ramakrishna, et~al.
\newblock Double-blind consent-driven data sharing on blockchain.
\newblock In {\em 2018 IEEE International Conference on Cloud Engineering
  (IC2E)}, pages 385--391. IEEE, 2018.

\bibitem{androulaki2018hyperledger}
Elli Androulaki, Artem Barger, Vita Bortnikov, Christian Cachin, Konstantinos
  Christidis, Angelo De~Caro, David Enyeart, Christopher Ferris, Gennady
  Laventman, Yacov Manevich, et~al.
\newblock Hyperledger fabric: a distributed operating system for permissioned
  blockchains.
\newblock In {\em Proceedings of the Thirteenth EuroSys Conference}, pages
  1--15, 2018.

\bibitem{rantos2018advocate}
Konstantinos Rantos, George Drosatos, Konstantinos Demertzis, Christos
  Ilioudis, Alexandros Papanikolaou, and Antonios Kritsas.
\newblock Advocate: a consent management platform for personal data processing
  in the iot using blockchain technology.
\newblock In {\em International Conference on Security for Information
  Technology and Communications}, pages 300--313. Springer, 2018.

\bibitem{ISO-24760}
Iso/iec 24760:2019 it security and privacy — a framework for identity
  management.
\newblock Technical report, New York, 1994.

\bibitem{eIDAS}
European Union.
\newblock {eIDAS} -- {Electronic Identification, Authentication and Trust
  Services}, {EU Regulation} 910/2014.

\bibitem{MUHLE201880}
Alexander Mühle, Andreas Grüner, Tatiana Gayvoronskaya, and Christoph Meinel.
\newblock A survey on essential components of a self-sovereign identity.
\newblock {\em Computer Science Review}, 30:80 -- 86, 2018.

\bibitem{w3cDID1.0}
D~Reed and M~Sporny.
\newblock {W3C} {Decentralized Identifiers} ({DIDs}) v1.0, 2017.

\bibitem{w3cVCDM1.0}
M~Sporny, D~Longley, and Chadwick D.
\newblock {W3C} {Verifiable Credentials Data Model} 1.0: Expressing verifiable
  information on the web, 2018.

\bibitem{asset}
{Algorand Official Documentation}, exploring features: Assets.
\newblock \url{https://developer.algorand.org/docs/features/asa/}, Last
  accessed May 2020.

\bibitem{victor2019measuringErc20}
Friedhelm Victor and Bianca~Katharina L{\"u}ders.
\newblock Measuring ethereum-based {ERC20} token networks.
\newblock In {\em International Conference on Financial Cryptography and Data
  Security}, pages 113--129. Springer, 2019.

\bibitem{PynaclDocu}
Pynacl - public key encryption.
\newblock \url{https://pynacl.readthedocs.io/en/latest/public/}, Last accessed
  May 2020.

\bibitem{NaClPaper}
Daniel~J. Bernstein.
\newblock Cryptography in nacl.
\newblock {\em Networking and Cryptography library}, 3:385, 2009.

\bibitem{rfc7748}
Adam Langley, Mike Hamburg, and Sean Turner.
\newblock {Elliptic Curves for Security}.
\newblock RFC 7748, January 2016.

\bibitem{poc}
Decentralized robinson list.
\newblock
  \url{https://gitlab.com/amauro/fob_robinson_list/-/tree/journal_version},
  Last accessed May 2021.

\bibitem{algorand}
{Algorand}: The first pure proof of stake blockchain platform.
\newblock \url{https://www.algorand.com/}, Last accessed May 2020.

\bibitem{ethfaqsharding}
Ethereum wiki project.
\newblock On sharding blockchains.
\newblock \url{https://eth.wiki/sharding/Sharding-FAQs} [Accessed 2021-05-05].

\bibitem{10.1145/3410699.3413800}
Gianmaria~Del Monte, Diego Pennino, and Maurizio Pizzonia.
\newblock Scaling blockchains without giving up decentralization and security:
  A solution to the blockchain scalability trilemma.
\newblock In {\em Proceedings of the 3rd Workshop on Cryptocurrencies and
  Blockchains for Distributed Systems}, CryBlock '20, page 71–76, New York,
  NY, USA, 2020. Association for Computing Machinery.

\bibitem{account}
{Algorand Official Documentation}, exploring features: Accounts.
\newblock \url{https://developer.algorand.org/docs/features/accounts/create/},
  Last accessed May 2021.

\bibitem{voigt2017eu}
Paul Voigt and Axel Von~dem Bussche.
\newblock The eu general data protection regulation (gdpr).
\newblock {\em A Practical Guide, 1st Ed., Cham: Springer International
  Publishing}, 2017.

\bibitem{binding}
Diego Pennino, Maurizio Pizzonia, Andrea Vitaletti, and Marco Zecchini.
\newblock Binding of endpoints to identifiers by on-chain proofs.
\newblock In {\em 2020 IEEE Symposium on Computers and Communications (ISCC)},
  pages 1--6, 2020.

\end{thebibliography}
